# Theoretical Model of Microparticle-Assisted Super-Resolution Microscopy

**A. R. Bekirov,* B. S. Luk'yanchuk, N. A. Lystseva, N.V. Grednev and A. A. Fedyanin**

*Faculty of Physics, Lomonosov Moscow State University, Moscow 119991, Russia*

*†The authors contributed equally to this Letter.*

** bekirovar@my.msu.ru*

**We present the three-dimensional theoretical model of microparticle-assisted super-resolution imaging that enables accurate simulation of virtual image formation. The model shows that partial spatial coherence of the illumination is a key prerequisite for achieving super-resolution. Rapid oscillations in the spatial correlation function of the radiation are demonstrated to significantly enhance the resolving capability of the microsphere, while the microsphere itself can locally increase the effective numerical aperture of the optical system up to unity. We further show that the optical resolution improves with increasing size of the investigated objects. An analytical estimate of the resolution criterion in microsphere-assisted imaging is provided. Overall, the results establish a consistent wave-optical framework that reproduces experimentally observed subwavelength imaging and clarifies the underlying physical mechanisms.**

**Introduction**. In 2011, a super-resolution imaging effect was discovered using dielectric microparticles [1], whereby placing a transparent microparticle of a few microns in diameter onto the sample enables the visualization of fine details that are unresolvable under conventional microscopy. This phenomenon arises due to the formation of a magnified virtual image by the microparticle. Since then, the technique has rapidly evolved, with efforts focused on improving image quality and controlling particle positioning for the visualization of extended structures. Recent experimental demonstrations have shown that combining microsphere-assisted microscopy with evanescent wave illumination [2] enables the imaging of low-contrast objects. This technique offers significant potential for the high-resolution study of biological specimens, a field where non-invasive methods are of paramount importance. A comprehensive review of theoretical and experimental studies can be found in Ref. [3].

Despite growing research interest, there is still no widely accepted theoretical model for this effect. Such a model is required to clarify the physical origin of the effect and to enable predictive optimization of imaging performance under different illumination and sample conditions. Many prior studies attempted to simulate radiation propagation and image formation in the presence of a microparticle, but due to computational complexity, most models are limited to two-dimensional configurations [4-6]. However, 2D models fail to capture the geometry of realistic samples—such as nanodisk arrays or patterned surfaces—which makes quantitative comparison with experimental data difficult, despite its importance for model validation.

In this work, we propose a full three-dimensional model for microparticle-assisted super-resolution imaging. The model accounts for the spatial coherence of illumination, broadband and narrowband spectra, and different illumination geometries. We validate the results against several experimental datasets. While a spherical microparticle is used in the simulations, the approach can be extended to other particle geometries.

**Image Field Formation Algorithm.** Let a spatially coherent incident field $\mathbf{E}_{inc}$ illuminate a given system at frequency $\omega$. As a result of diffraction, a new field distribution E is formed in the surrounding space. We assume that this field E is observed by an ideal aberration-free optical system with 1:1 magnification, located at infinity, in the half-space z>0. The resulting image field $\mathbf{E}_{im}$ can be calculated using the following expression:

$$\mathbf{E}_{im}=\iint_{\Gamma}\left(\frac{i\mu_0}{n_0}\left[\mathbf{n},\mathbf{H}^*\right]\left(\bullet G+\left[\bullet,\nabla G\right]\right)-\left[\left[\mathbf{n},\mathbf{E}^*\right],\nabla G\right]\right)\frac{k^2 dS}{4\pi}, \quad (1)$$

where $G=\exp(ik|r-r_0|)/k|r-r_0|$, $\nabla=k^{-1}\partial/\partial \boldsymbol{r}_0$, $k=2\pi/\lambda$ and $\Gamma$ is the surface homotopic to the plane, $\mu_0$ and $n_0$ represent the magnetic permeability and the refractive index of the surrounding medium, respectively. In our calculations, we assumed $\mu_0=n_0=1$. To compute the image produced by an incoherent source, the image field must be intensity-averaged over different realizations of the incident fields $\mathbf{E}_{inc}$. Additionally, if the source has a broadband spectrum, averaging over frequency is also required, so that:

$$I_{im}^{non.coh}(r)=\frac{c}{8\pi}\iint\left|\mathbf{E}_{im}\left(r,\omega,\mathbf{E}_{inc}\right)\right|^2 dE_{inc}d\omega. \quad (2)$$

The differential element $dE_{inc}$ parameterizes the geometry of the incident illumination. The specific form of the field $\mathbf{E}_{inc}$ depends on the illumination conditions of the system. In the most commonly used Köhler illumination scheme, the incident field is given by $\mathbf{E}_{inc}=\mathbf{E}_0\exp(i\mathbf{k}\cdot\mathbf{r})$, with $dE_{inc}=dk_x dk_y$, i.e., the sample is illuminated by plane waves coming from different angles. In the case of critical illumination, the incident field is a focused Gaussian beam. In the coordinate system where the propagation direction of the beam aligns with the *z*-axis, the field in the focal plane is given by $\mathbf{E}_{inc}=\mathbf{E}_0\exp(-(x^2+y^2)/(2w_0^2))$, where again $dE_{inc}=dk_x dk_y$, with $k_x$, $k_y$ characterizes the angle of incidence. The general calculation scheme is illustrated in Fig. 1.

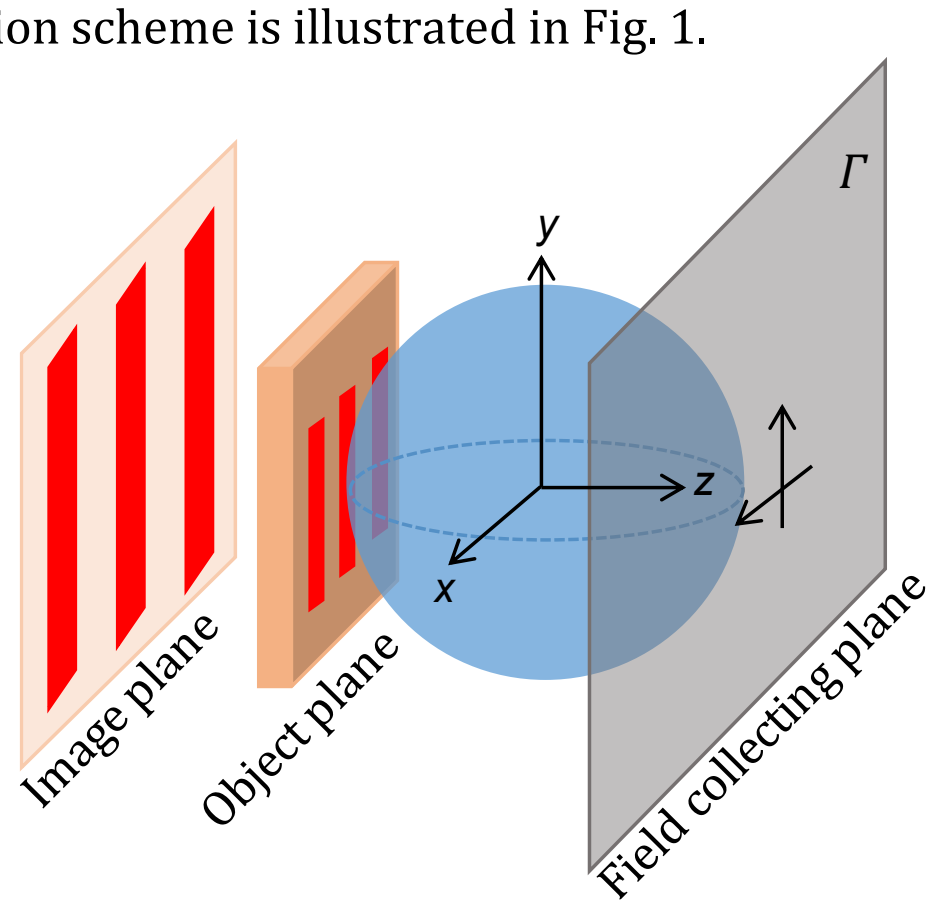

Fig. 1. General simulation scheme.

**Diffraction field models for the microsphere.**

To implement the proposed algorithm, it is necessary to determine the electric and magnetic fields **E** and **H** on the specified surface Γ above the microsphere (see Fig. 1). In this work, we employed two approaches for calculating these fields. The first approach, referred to as the Full Model (FM), involves a complete simulation of the electromagnetic field in the volume surrounding the microsphere. The second approach, termed the Simple Model (SM), considers only the field propagation in the vicinity of the contact region between the microsphere and the sample, which significantly reduces the computational domain.

***A. Full model simulation.*** The simulation of the fields **E**, **H** was carried out using FDTD algorithm on a standard square computational grid with a spatial step size of $dx=dy=dz=0.027$ µm, and a time step of $dt=0.99\cdot dr/c=0.052$ fs, where *c* is the speed of light. In the vicinity of the structures under consideration, the spatial grid was locally refined in order to accurately resolve the geometrical features of the objects. The simulation domain was approximately 9 µm×9 µm×6.8 µm. The lateral size of the surface *Γ* sets the maximum collection angles from the sample, defining the effective objective numerical aperture $NA_{obj}$. In the present model, $NA_{obj}\approx0.7$. All materials are considered non-dispersive with a constant refractive index, the microsphere had a diameter of 5 µm. Fig. 1 shows the general layout of the simulation. Standard built-in sources from the Lumerical FDTD package were used to generate plane wave and Gaussian beam excitations. The field $\mathbf{E}_0$ was assumed to be polarized either as a TM or TE wave with respect to the substrate surface. The surface Γ was implemented as a monitor placed directly above the microsphere, performing spectral Fourier transforms for a given set of frequencies. This provided the field components $\mathbf{E}(x, y, \omega_i)$ and $\mathbf{H}(x, y, \omega_i)$ which were then used in equation (1).

***B. Simple model simulation.*** In the SM, the fields **E** and **H** are simulated only in the contact region between the microsphere and the sample. Knowing the fields in the sample plane, the subsequent propagation of light through the microsphere can be analytically calculated using the Huygens–Fresnel principle in combination with Mie theory. This approach neglects the secondary illumination of the sample by the waves circulating inside the microsphere, which may have a significant impact under resonant conditions.

Depending on the visualization geometry, the source field in the SM must be defined differently. In the reflection mode, the microsphere affects the illumination field, since the radiation first passes through the microsphere and then reaches the sample. To account for this effect, we applied Mie theory for a sphere in free space to calculate the diffraction of the incident field on the microsphere and used the resulting near field distribution as the source field. Note that the field distribution within a particle placed on a substrate deviates from that in free space due to the interference of waves reflected at the boundaries. The contribution of these effects may become significant in the vicinity of resonances and can potentially lead to qualitative differences between the SM and FM descriptions. Nevertheless, for the non-resonant configurations examined in this work, their impact appears to be predominantly quantitative, resulting mainly in variations of the image amplitude, while no qualitative changes in the image formation were observed. In the transmission mode, where the illumination first passes through the sample and then through the microsphere, the source field can be defined in the same way as in the full model, without any modification of the incident wavefront. See the Appendix for details of the SM algorithm.

***C. Comparison of the models.*** Fig. 2 shows a comparison of the simulation domains in the two models, as well as the positions of the field monitors and the source regions within the computational volume. In the SM, the fields on the surface Γ are calculated using the Huygens–Fresnel principle applied to the fields defined on the auxiliary surface $\Gamma_{Near}$ (see Fig. 2(d)).

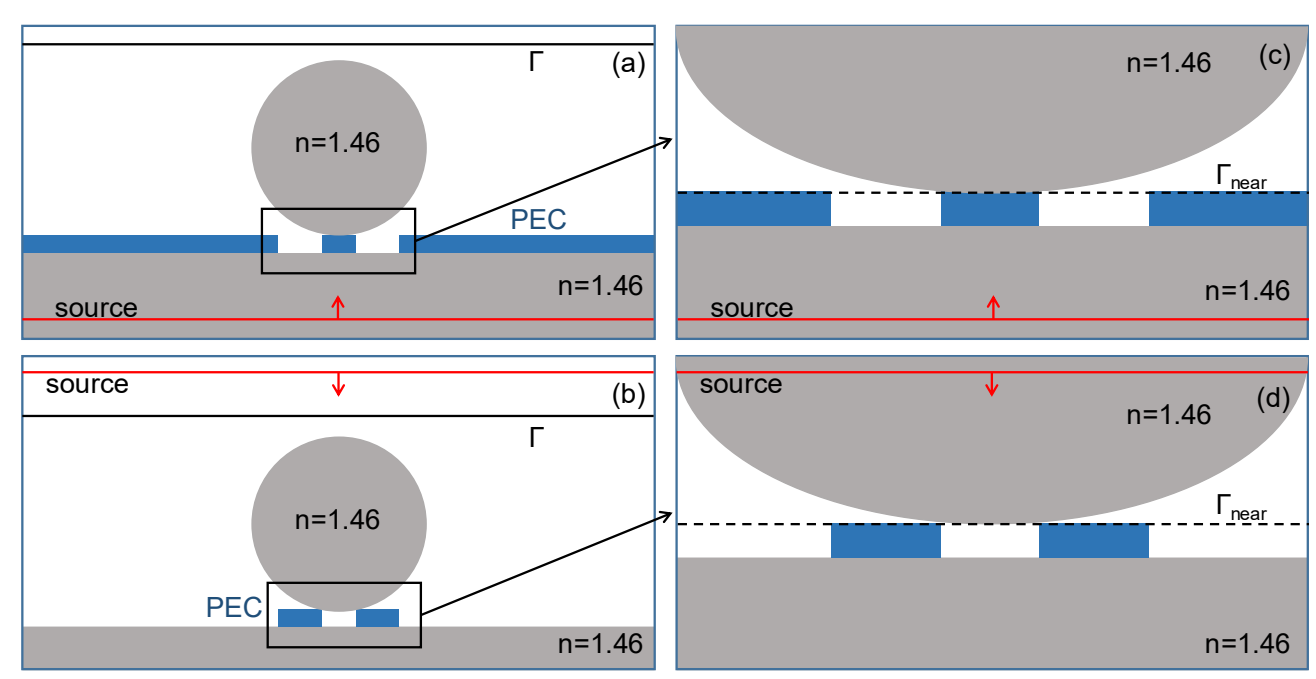

Fig. 2. Comparison of the simulation domains (not to scale) for the FM and SM. (a–b) Side views of the FM in the reflection and transmission modes, respectively. (c–d) Corresponding configurations for the SM. In panel (d), the source field is defined with the wavefront distortion induced by the microsphere taken into account. The positions of the field monitors coincide with the surface *Γ*. The microsphere radius is 2.5 µm.

Two circular apertures with diameters of 250 nm were considered as test objects. The apertures were patterned in a perfectly conducting film of 10 nm thickness, with a center-to-center separation of 300 nm. The illumination wavelength was 500 nm. In the reflection mode, the structure was complementary to the transmissive one, with identical geometric parameters. The illumination was arranged in a Köhler configuration with a cone angle of π/4, corresponding to an illuminator numerical aperture of $NA_{ill} = \sin(\pi/4) \approx 0.7$. Under these conditions, the two apertures are unresolved in direct imaging (see Figs. 3(a–d)). However, when viewed through the microsphere, the apertures become distinguishable, as confirmed by both models in the transmission mode (Fig. 3(b–c)) and in the reflection mode (Fig. 3(f–g)).

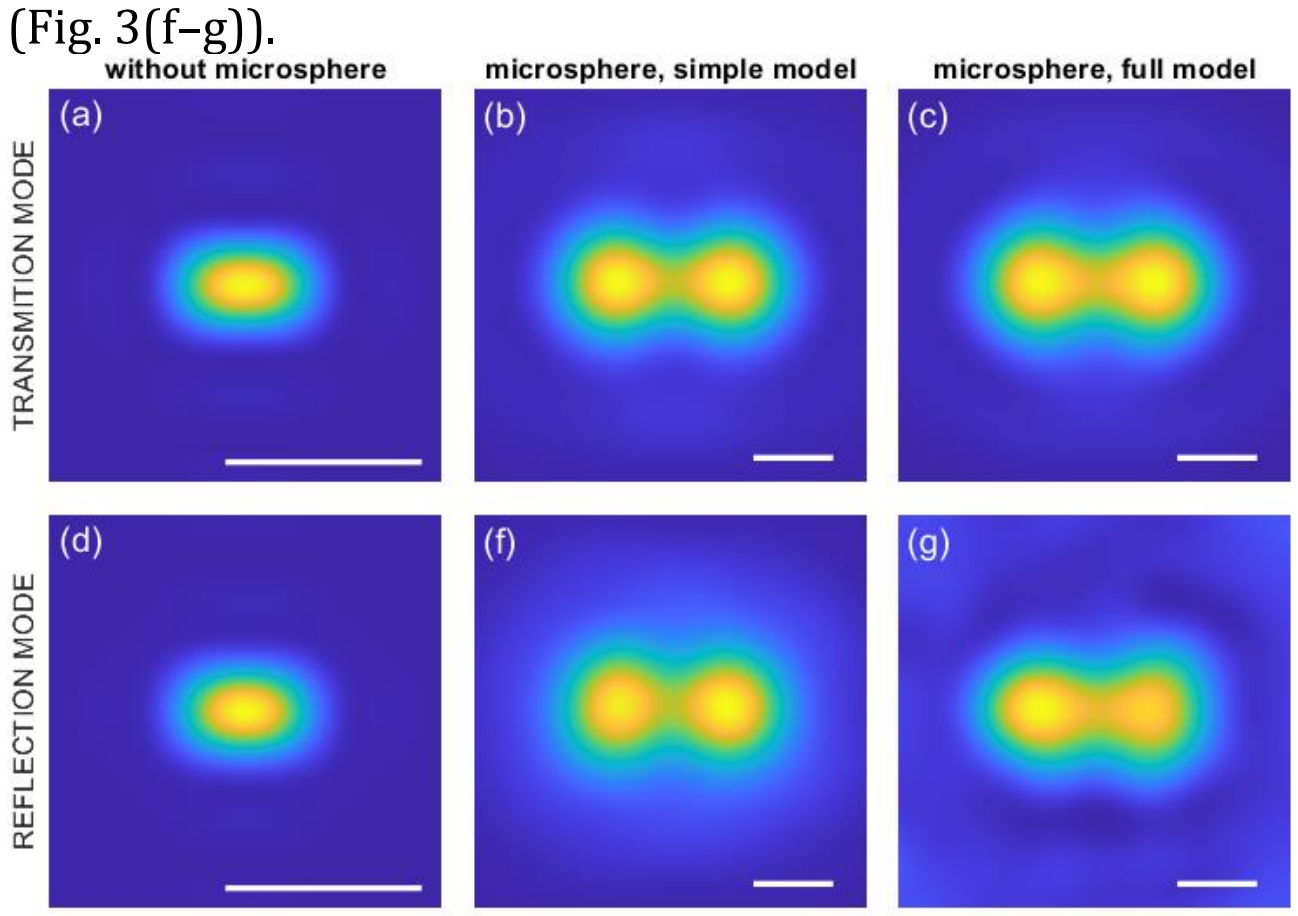

Fig. 3. Comparison of the FM and SM for the formation of the virtual image by a microsphere. Images of two nanoscale apertures with a diameter of 250 nm and edge-to-edge separation of 50 nm in a perfectly conducting film in reflection mode: (a) without the microsphere, (b) with the microsphere using the SM, and (c) using the FM. Panels (d–g) show the corresponding comparison for two disks in reflection mode. In free space, the image plane coincides with the object plane. When the microsphere is present, the image is observed in the plane z/R = –5 relative to the center of the microsphere. Scale bar: 1 μm. Ratio of peak intensities: (b)–(c) 1:1.07; (f)–(g) 1.15:1.

For the Simple Model, the simulation domain has approximate dimensions of 4×4×1.5 μm, whereas the Full Model employs a computational volume of 10×10×7 μm, respectively. Due to the smaller numerical domain, simulations using the SM run roughly 100 times faster than those with the FM.

## Full model result

***A. Broadband Source with Köhler Illumination Scheme.*** To validate our model, we begin by simulating the imaging of a Blu-ray disc surface, a benchmark structure that has been widely studied in experimental works [1,7]. The disc was modeled as a multilayer structure consisting of the following elements: an infinite glass substrate (refractive index n=1.46), a 30 nm thick metal layer (with refractive index n=0.77+*i*6.08), and a patterned surface composed of parallel trenches 200 nm wide, separated by 100 nm gaps and 30 nm deep, with the same refractive index as the metal layer. Let us consider the geometry proposed in the experimental work [7]: BTG microspheres with a refractive index of $n$ = 1.95 were immersed in an immersion liquid (water, $n$ = 1.33). A detailed illustration of the geometry used is provided in the Appendix.

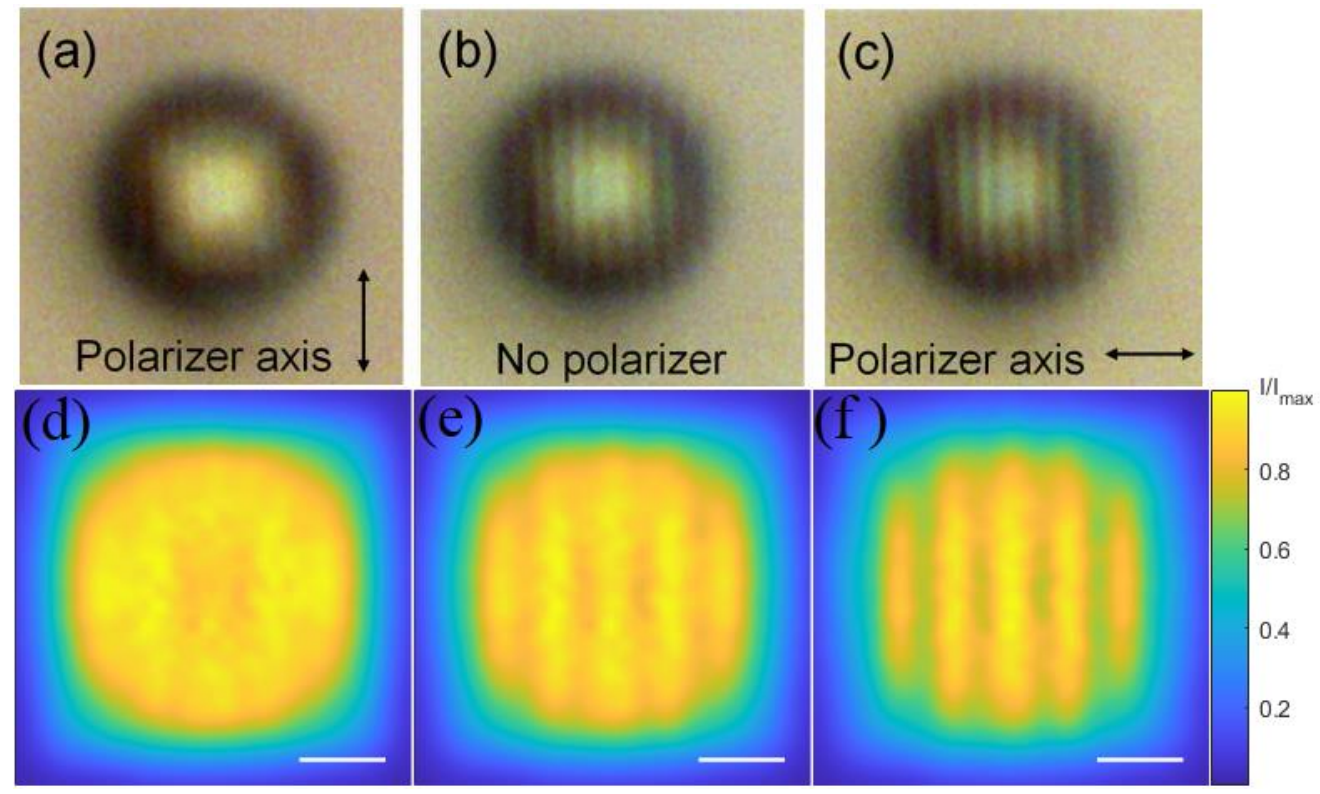

Fig. 4. (a)–(c) Virtual images of the Blu-ray disc surface obtained at different polarizer orientations, according to Ref. [6]. (d)–(f) simulation results. The observation plane for images (d)–(f) is located at a depth of $z$ = –6.25 μm below the substrate surface. Note that the microsphere sizes in images (a)–(d) and (d)–(f) are different. The relative intensities of the maxima in (d)–(f) are 1:1.8:1.1, respectively. Scale bar: 2.5 μm.

The simulation results are shown in Fig. 4. Similar to Ref. [7], a strong dependence of the observed pattern on the polarization of the recorded image is observed. To simulate the image at different polarizer orientations, the calculation of the image intensity (2) included only the $x$- or $y$-component of the field $\mathbf{E}_{im}$ (1).

The illumination cone had an angular aperture of π/9, and included 50 uniformly distributed plane-wave directions. The source spectrum had a peak at $\lambda$ = 500 nm and spanned wavelengths from 400 to 700 nm (for more details, see the Appendix). The illumination geometry was manually adjusted to achieve the best agreement between the simulated and experimental results reported in Ref. [7]. A strong dependence of the image contrast and pattern on the illumination geometry was observed.

The presented model also allows the study of visualization in transmission mode. To validate it, we simulated the results from Ref. [1], where the visualization of slits in a metallic screen was considered. The sample consists of a 30 nm-thick metal film deposited on a glass substrate, with four 360 nm-wide slits spaced 130 nm apart, matching the experimental configuration used in Ref. [1]. The refractive index of the metal, n = 3.2+*i*3.3, corresponds to chromium (Cr). The sample was illuminated in transmission mode using a Köhler illumination scheme, within a half-angle of π/4, uniformly sampling 50 plane-wave directions within this cone. The simulation results are shown in Fig. 5.

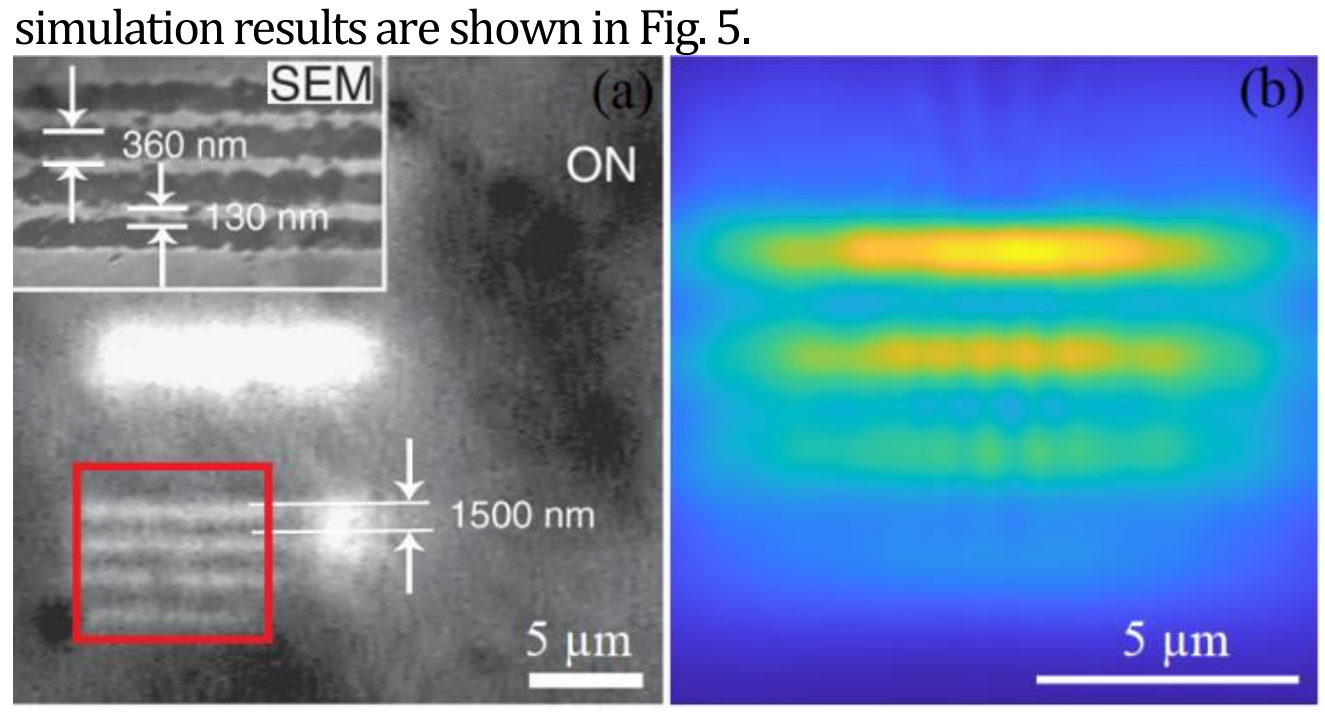

Fig. 5. Comparison of theoretical model results and experimental data. (a) Experimental observation of an array of four slits in a metallic screen. The inset shows an SEM image of the sample with dimensions

corresponding to those used in the theoretical model. (b) Simulation of the image in a plane located 7 μm below the substrate surface. The modeled area is marked with a red square.

***B. Monochromatic Source at 405 nm with Critical Illumination.*** Among the most notable achievements of microsphere-assisted imaging is its integration with confocal microscopy which has demonstrated record-breaking resolution of nanostructures. In particular, nanodisks with diameters of 135 nm spaced only 25 nm apart were successfully resolved [8].

Unlike conventional microscopes operating under Köhler illumination, a confocal microscope uses a focused beam to illuminate each point of the sample individually. The image is formed by scanning the region of interest and reconstructing the complete image from individual focus points. Therefore, to model image formation in a confocal microscope, it is necessary to use a focused beam rather than plane waves.

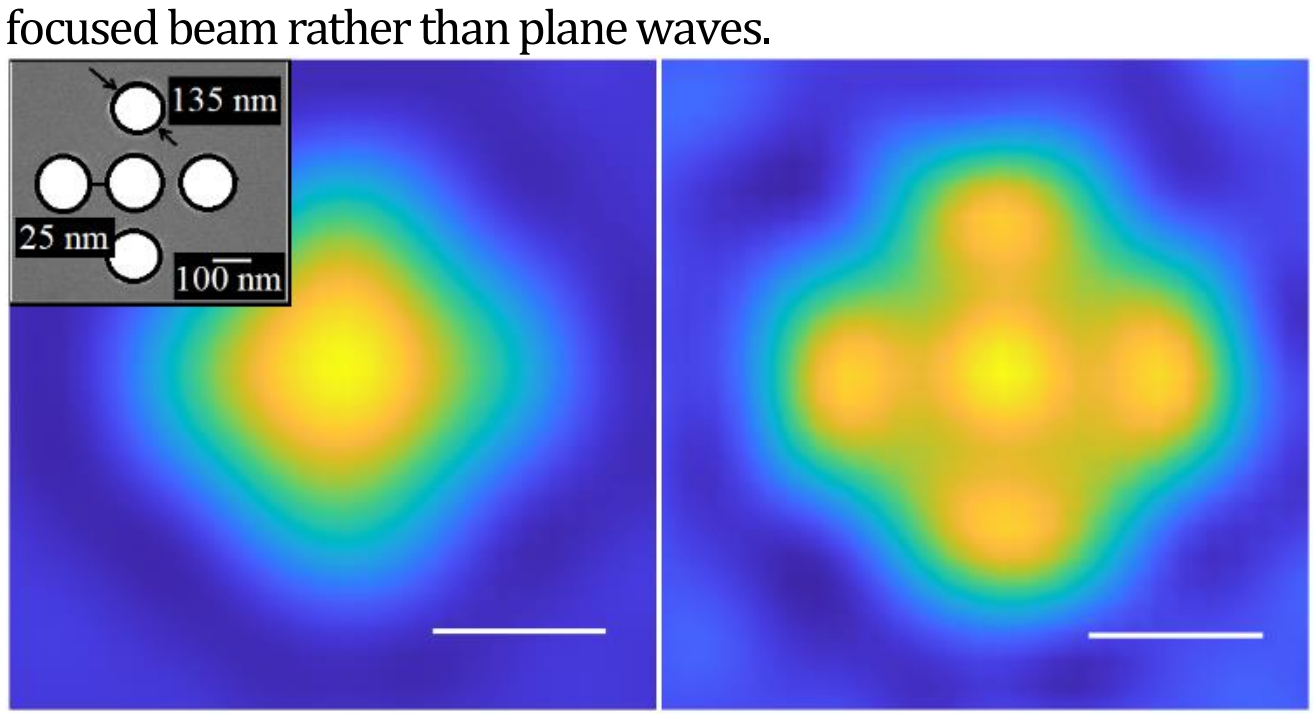

Fig. 6. (a) Virtual image of five nanodisks with the geometry shown in the inset, obtained under critical illumination. (b) Corresponding image with the illumination suppressed inside the cone $\theta < \pi/12$. Both images are taken in the focal plane at $z = -10$ μm below the substrate surface. Scale bar: 1 μm.

In our simulations, we assumed that the source is focused on the sample surface, thereby implementing the classical critical illumination scheme. The illumination cone had an angular aperture of $\pi/4$ and consisted of 103 uniformly distributed wave directions. It is worth noting that achieving super-resolution in this configuration requires a sufficiently wide illumination cone. The disks were positioned on a transparent substrate (n = 1.46), coaxially aligned with the center of the microparticle.

In order to achieve optimal image contrast under the given illumination conditions, it is essential to suppress radiation propagating close to the surface normal. To this end, the signal amplitude of rays focused within the illumination cone at angles $\theta < \pi/12$ was attenuated by a factor of 10. In practice, this requirement can be readily fulfilled by blocking the light originating from the central part of the source. This step is crucial to avoid dominance of the central lobe, which would otherwise obscure finer image details. The results are presented in Fig. 6. We also applied our model to simulate an array of perfectly conducting dimers with 120 nm width and 60 nm spacing, as described in Ref. [9]. In this case as well, the structure could be successfully visualized using our approach (see the Appendix).

These examples enable a fundamental study of the resolving power of a microsphere and the key factors underlying its super-resolution. In our earlier work [10], based on a two-dimensional numerical model, we showed that as the size of the objects imaged through a microsphere decreases, the resolution approaches the classical Abbe limit of $\lambda/2$.

To extend this hypothesis to three dimensions, we modeled each nanodisk as a point-like incoherent source placed at its center. The simulations demonstrated that such sources are indistinguishable. A similar calculation, with the sources located on the nanodisk surfaces, yielded the same result: the structures remained unresolved. Although the disks retain their characteristic size, each of them emits independently. These findings highlight a fundamental requirement for super-resolution: spatial coherence of illumination.

The presence of partial spatial coherence effects in the illumination indicates that the simulation results strongly depend on the specific field distribution in the region of the disks. The substrate material can also influence this distribution, as noted in Ref. [11]. Our model shows that if a similar calculation is performed without accounting for the substrate, i.e., considering the disks in free space with the microparticle above them, the structure becomes indistinguishable, even accounting for the spatial coherence of the illumination.

## Simple model result

The FM enables the investigation of a wide range of problems related to microsphere-based imaging; however, this approach requires a significant computational effort. For a fixed system geometry, the FM may be appropriate, but when exploring the influence of variations in geometry or illumination, the computations become prohibitively complex. For this reason, in the present work, we employ simulations according to the SM with the illumination according to the Köhler scheme at a wavelength of $\lambda = 500$ nm.

***B. Influence of spatial coherence effects.*** As shown for nanodimers, partial-coherence effects in microsphere-based imaging can enhance resolution, where the objects no longer behave as independent radiators. Here, we consider off-axis annular or dark-field illumination to increase the coherence of the radiation. This illumination scheme enhances the resolving power of a microparticle, a concept first introduced in [12] for a two-dimensional model. Fig. 7 presents transmission-mode images of an array of nine apertures in an ideal conductor, each with a diameter of 180 nm and a period of 200 nm, obtained for different illumination geometries. The imaging plane is located at $z/R=-5$ with respect to the center of the microparticle. The simulations indicate that the virtual image forms near the geometrical focus at $z/R \approx -3$ and improves upon further downward shift of the imaging plane, while excessive displacement leads to image degradation.

In this example, the distance between adjacent apertures is below the diffraction limit. Under conventional Köhler illumination, the structure cannot be resolved by the microparticle (Fig. 7(a)). Suppression of the low spatial-frequency components of the source increases the spatial coherence of the incident field, leading to improved resolution (Fig. 7 (b)-(c)). The peak intensity amplitudes in Fig. 7(a)–(c) are in the ratio 24.8:7.4:1.9. After normalization of the incident field by the number of plane-wave components used in the calculation, the corresponding ratio becomes 8.2:3.2:1.6.

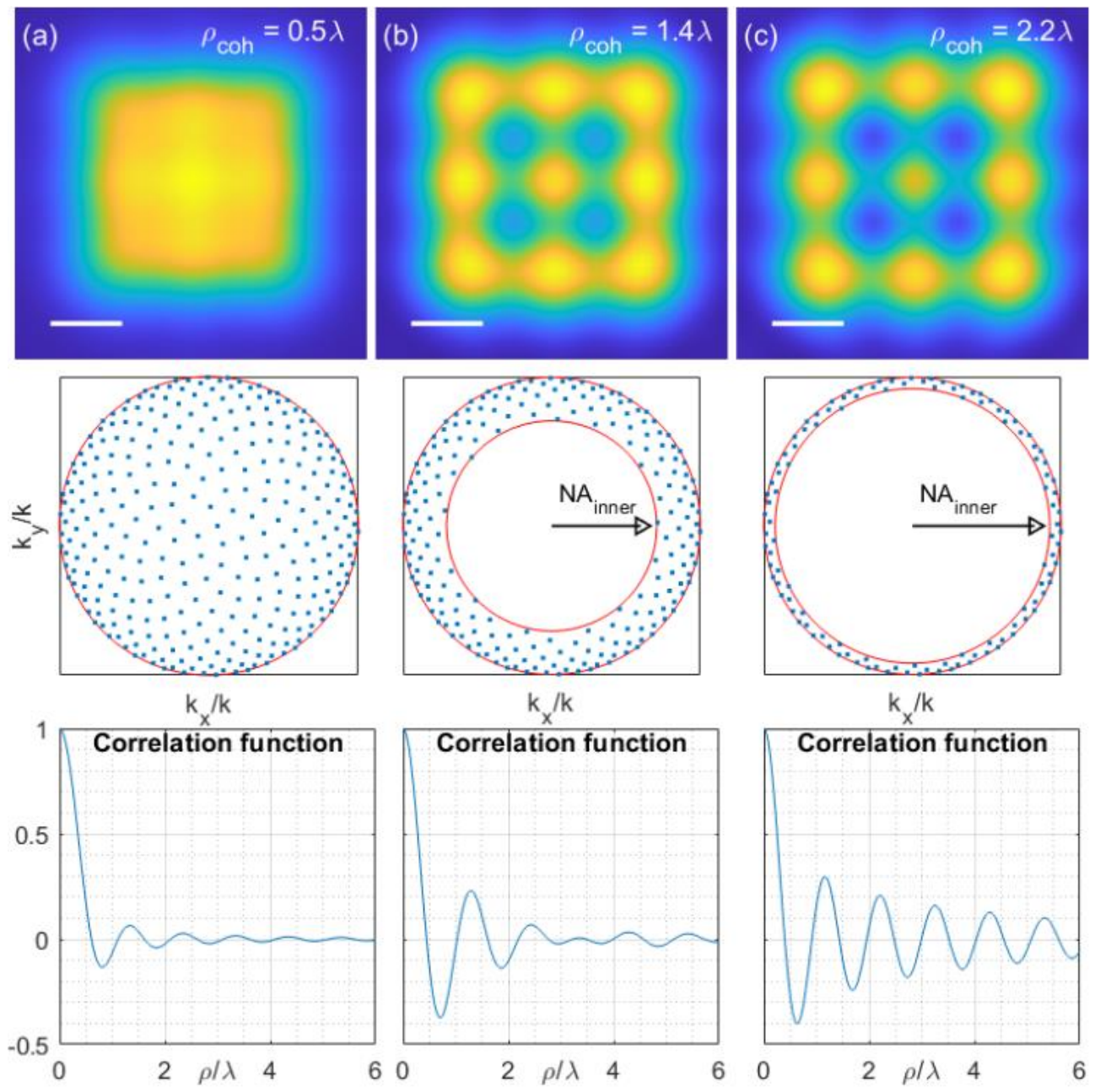

Fig. 7. Image field of an array of nine nanoapertures (diameter 180 nm, period 200 nm) in an ideally conducting film for different spatial coherence radii: (a–c) with 0.61λ, 1.6λ, and 3.5λ, respectively (λ = 500 nm). The illumination directions used in the simulations are shown at the bottom of each panel, together with the corresponding correlation functions. The dots indicate the plane-wave components included in the calculation. The focal plane is located at z/R=−5. Scale bar: 1 μm.

The enhancement of the resolving capability of the microparticle under annular illumination is observed for structures whose characteristic dimensions are comparable to the illumination wavelength. When the diameter of the nanoapertures is reduced to 40 nm (λ/12.5) while keeping the center-to-center separation unchanged, the structure becomes unresolvable (see Appendix). Partial coherence effects may also influence the resolution in free space; however, this structure remains unresolvable for all considered illumination schemes, even when the objective numerical aperture reaches $NA_{obj} = 1$ (see Appendix).

Under annular illumination, the normalized correlation function [13] of the radiation is defined as

$$\gamma(\rho)=\frac{\left\langle\left(\mathbf{E}(r),\mathbf{E}^*(r+\rho)\right)\right\rangle_t}{\sqrt{\left\langle|\mathbf{E}(r)|^2\right\rangle\left\langle|\mathbf{E}(r+\rho)|^2\right\rangle}}=\frac{\int_{NA_{inner}}^{NA_{outer}} xJ_0(k\rho x)dx}{\frac{k^2}{2}\left(NA_{outer}^2-NA_{inner}^2\right)}$$

where $NA_{inner}$ and $NA_{outer}$ specify the lower and upper bounds of the transverse wave vectors. The spatial coherence of the field is characterized by the spatial coherence radius $\rho_{coh}$, i.e., the distance over which a significant degree of mutual coherence is preserved. Conventionally, the spatial coherence radius $\rho_{coh}$ is defined by the first zero of the function $\gamma$. In this work, $\rho_{coh}$ is defined as the distance $\rho$ beyond which the maximum value of $|\gamma|$ does not exceed 0.2, which accounts for possible oscillations of $\gamma$ beyond its first zero.

For annular illumination, $\gamma$ exhibits pronounced oscillations, in contrast to uniform (filled-cone) illumination with the same $\gamma$, which leads to an enhanced resolving capability of the microparticle in the virtual image. It should be emphasized that this improvement is governed by partial coherence effects: for fully coherent illumination ($\rho_{coh}\to\infty$), the structure remains unresolvable. At the opposite limit, $\rho_{coh}\to 0$, which is unattainable in the classical Köhler illumination scheme, no enhancement of the resolving capability is observed either. Such radiation can only be realized when the source is placed in close proximity to the sample, for example, beneath the nanoscale holes.

The enhancement of the resolving capability in microsphere-based imaging can be demonstrated analytically using a model of point sources, without resorting to numerical simulations. Let's consider an array of N point dipoles with partially correlated random phases $\varphi_j$. The total electromagnetic field generated by the system is described by

$$\Delta\mathbf{E}+k^2\mathbf{E}=-4\pi k^{-1}\sum_{j=1}^{N}e^{i\varphi_j}\delta(\mathbf{r}-\mathbf{r}_j)\mathbf{e}_x\ ,$$

where the statistical properties of the phases are governed by the correlation function

$$\left\langle e^{i(\varphi_m-\varphi_m)}\right\rangle=\gamma\left(|\mathbf{r}_m-\mathbf{r}_n|\right).$$

The total intensity of the virtual image is expressed in terms of the individual image fields $\boldsymbol{E}_{im}^{(j)}$ as

$$I_{im}=\frac{c}{8\pi}\left\langle|\mathbf{E}_{im}|^2\right\rangle\sim\sum_{i,j=1}^{N}\left(\mathbf{E}_{im}^{(i)},\mathbf{E}_{im}^{(j)*}\right)\gamma\left(|\mathbf{r}_m-\mathbf{r}_n|\right).$$

The fields $\boldsymbol{E}_{im}^{(j)}$ are obtained analytically from Mie theory and Eq. (1). Three forms of the second-order correlation function $\gamma$ are analyzed: (i) fully incoherent dipoles; (ii) partially coherent dipoles with a correlation function corresponding to Köhler illumination with $NA_{ill}$ = 1; and (iii) annular illumination with $NA_{inner} = NA_{outer} = 1$, corresponding to ring illumination with $k_z = 0$. The corresponding correlation functions are given by

$$\gamma(\rho)=\begin{cases}\delta(k\rho)\\ 2J_1(k\rho)/(k\rho)\\ J_0(k\rho)\end{cases},$$

where $\delta$ is the Kronecker delta and $J_n$ denotes the Bessel function of the first kind.

The source array consists of two square regions of size w=λ/3, separated by a distance d=0.185λ, located in the plane z=−R and arranged symmetrically with respect to the center of the microsphere. The virtual image is observed through an objective with numerical aperture $NA_{obj} = 0.9$.

Fig. 8 shows the resulting virtual images in free space and in the presence of a microsphere (n = 1.46, 2πR/λ = 30) for different forms of $\gamma$. The comparison reveals a pronounced dependence of the microsphere-induced resolution enhancement on the shape of the correlation function.

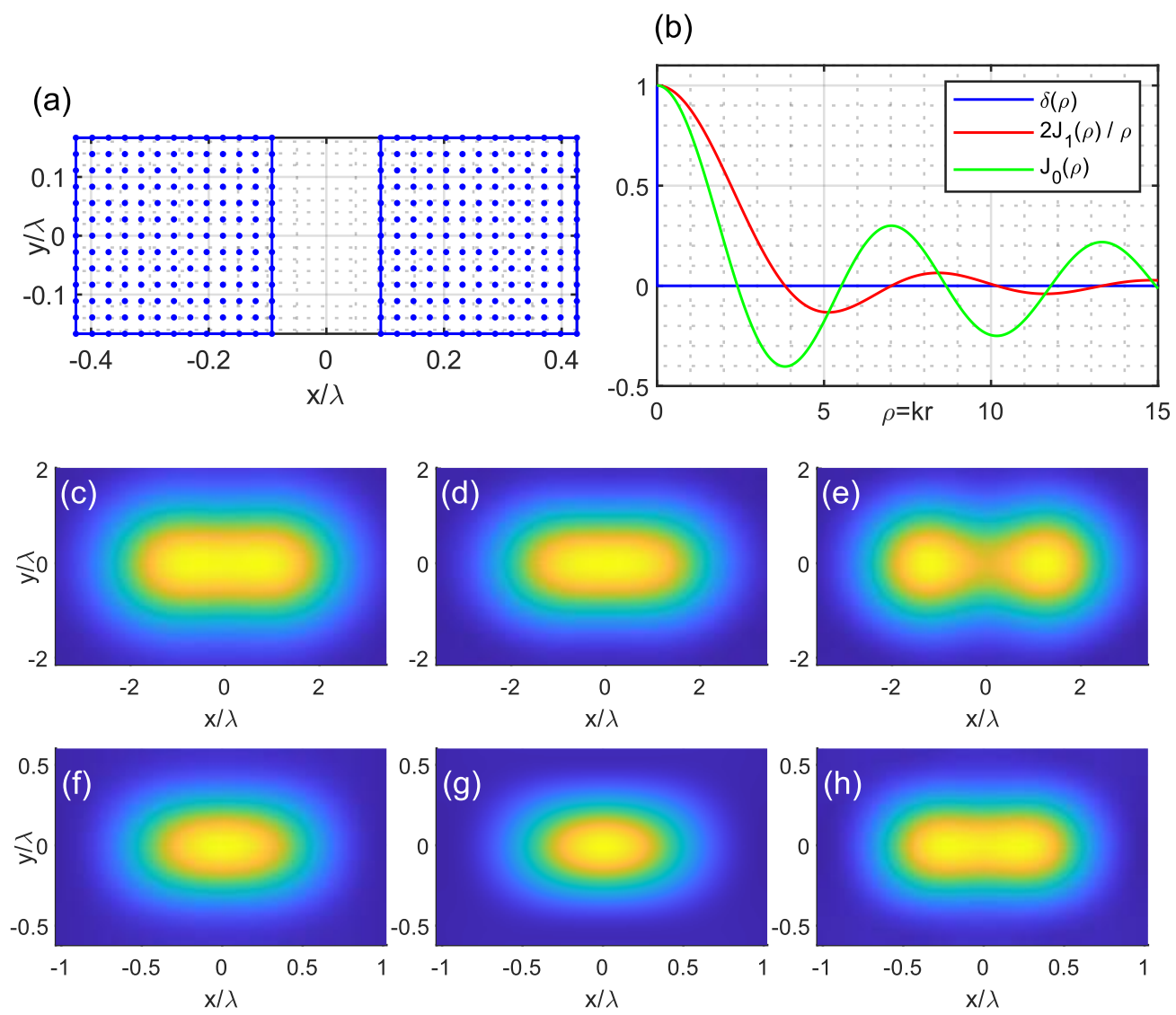

Fig. 8 (a) Point-sources geometry. (b) Correlation functions used in the calculations. (c–e) Virtual images of the dipole array formed by the microparticle (n=1.46, 2πR/λ=30) in the plane z/R=−4.5 using $\gamma = \delta(\rho)$, $2J_1(\rho)/\rho$, $J_0(\rho)$. (f–h) Corresponding free-space images; the image plane coincides with the source plane.

For sufficiently large separations *d*, similar coherence-induced effects can also be observed in free space. Therefore, the microsphere primarily amplifies the influence of partial coherence on the resolving capability, rather than introducing a fundamentally new resolution mechanism.

***B. Resolution analysis.*** In this section, we examine the resolution limit of the microsphere as the size of the investigated structures is progressively reduced. We consider standard Köhler illumination in the transmission configuration, as well as the case where the central part of the illumination within the inner numerical aperture, $NA_{inner}$, is blocked (Fig. 7(b, c)). The test objects are identical to those shown in Fig. 3(b–f) and consist of two disks in a PEC screen placed on a dielectric substrate. The image plane is located at $z/R = -5$ relative to the center of the microsphere. The surface $\Gamma$ is defined as a square region of size 4Rx4R at $z/R = 1.1$, which corresponds to an objective numerical aperture of $NA_{obj} \approx 0.72$. The numerical aperture of the illuminator is chosen as $NA_{ill}$= sin(3*pi/8). Note that, since the rays transmitted through the microparticle are confined within a limited divergence cone, increasing the size of plane $\Gamma$ - and thus the objective numerical aperture $NA_{obj}$ - results in only minor modifications of the image formed by the microparticle. This trend is consistent with the experimental observations reported in Ref. [14].

The optical resolution is defined as the minimum edge-to-edge distance between the structures at which they remain distinguishable according to the Rayleigh criterion, i.e., when the intensity dip between adjacent maxima reaches 80%. The resolution was determined with an accuracy of λ/50 using the same algorithm as described in Ref. [10]. Figure 9 shows the dependence of the optical resolution on the structure diameter. Based on these calculations and in agreement with two-dimensional model [10], the resolution limit of the microparticle under Köhler illumination can be expressed as the following criterion:

$$d + w \approx \frac{\lambda}{2}, \qquad (3)$$

where *w* is the characteristic object size and *d* is the minimum edge-to-edge separation. Eq. (3) holds for $w < \lambda/2$. A similar scaling is observed in free space, i.e., without the microparticle.

$$d + w \approx 1.22\frac{\lambda}{2NA_{obj}} \qquad (4)$$

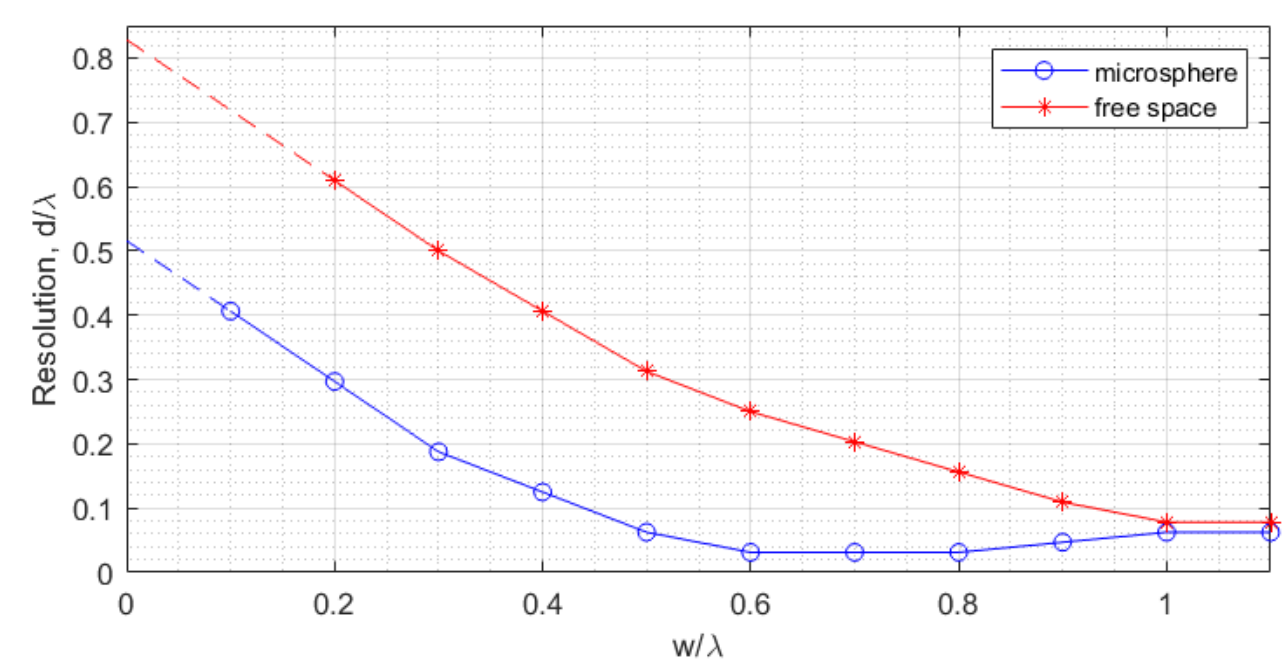

Fig. 9. Optical resolution as a function of the nanohole diameter under Köhler illumination ($NA_{ill}\approx 0.92$, $NA_{obj}\approx 0.72$) for microsphere-assisted imaging and free-space imaging. The dashed lines indicate extrapolated behavior in the limit of vanishing nanohole diameter, where direct numerical calculations become impractical due to the required refinement of the computational grid near the apertures.

From relations (3) and (4), it follows that under classical Köhler illumination, isolated objects in the form of nanoholes cannot be resolved by the microsphere beyond the fundamental diffraction limit, since the center-to-center separation remains close to λ/2. A comparison of relations (3) and (4) indicates that, in the considered configuration, the resolution enhancement provided by the microsphere relative to free space is primarily due to a local increase in the effective objective numerical aperture, as it is noted in Ref. [15]. However, NA is bounded by a limiting value of unity; consequently, this effect alone cannot account for the resolution of objects with a center-to-center separation smaller than λ/2.

For hole diameters exceeding λ/2, the optical resolution in free space and in the presence of the microsphere becomes nearly identical. In this regime, the objects appear as two touching ring-like intensity distributions and remain distinguishable even at zero edge-to-edge separation. Therefore, a meaningful comparison of resolving power for $w > \lambda/2$ requires objects of a different geometry, such as square apertures.

Under annular illumination, the microsphere enables resolution at separations below λ/2. We fix the center-to-center distance at $d+w=0.4\lambda$ and characterize image quality by the contrast $C=I_{middle}/I_{max}$, defined as the ratio of the intensity at the midpoint between adjacent maxima to the maximum image intensity. At $C$=1, the objects are completely merged, while at $C$=0.8 they are just resolvable according to the Rayleigh criterion.
Fig. 10 shows the dependence of $I_{middle}/I_{max}$ on the inner numerical aperture $NA_{inner}$ for a fixed outer numerical aperture $NA_{outer} = 1$, both for the microsphere (Fig. 10(a)) and for free space (Fig. 10(b)). The comparison demonstrates that the microsphere strongly enhances partial-coherence effects under annular illumination.

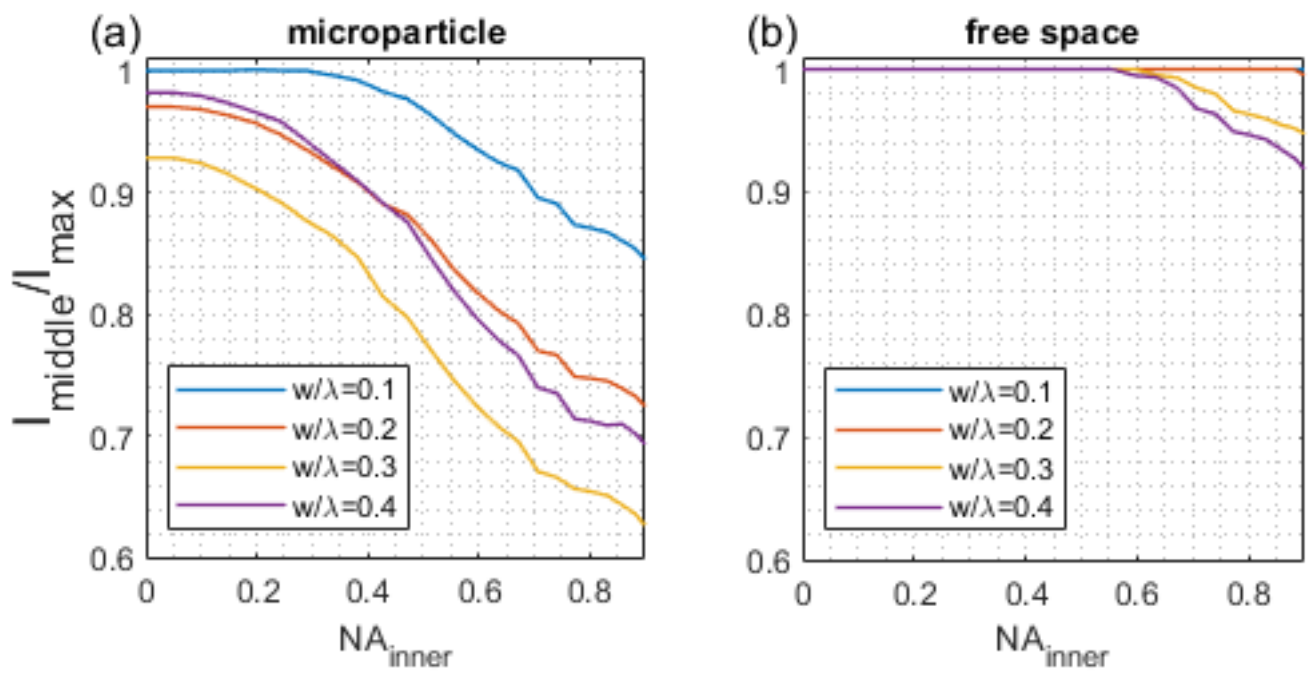

Fig. 10. Image contrast $C = I_{middle}/I_{max}$ for nanoholes of different diameters $w$ at a fixed center-to-center separation of $0.4\lambda$ as a function of the inner numerical aperture $NA_{inner}$: (a) microsphere-assisted imaging and (b) free-space imaging. As seen in (a), increasing the nanohole diameter $w$ leads to higher image contrast and enhanced resolving capability.

As shown in Fig. 10(a), increasing the nanohole diameter from $0.1\lambda$ to $0.3\lambda$ leads to a monotonic increase in image contrast, improving the resolving capability even for conventional Köhler illumination ($NA_{inner} = 0$). However, when $w = 0.4\lambda$, corresponding to touching nanodisks, the contrast drops sharply to $I_{middle}/I_{max} = 0.98$. The opposite trend was observed for the condition $d+w=0.3\lambda$. The contrast variation from C=1 ($NA_{inner}$=0) to C=0.75 ($NA_{inner}$=0.9) occurs exclusively at $w=0.3\lambda$, corresponding to the case of touching apertures. This behavior indicates that the resolving power of the microparticle in this imaging mode exhibits a complex nonlinear nature.

An experimental study [16] reported an improvement in optical resolution under annular illumination in free space. However, a direct comparison of Figs. 10(a) and 10(b) shows that increasing $NA_{inner}$ leads to a much stronger contrast improvement —and thus a more pronounced resolution enhancement—in the presence of a microsphere. This enhancement is particularly pronounced for structures with larger diameters.

## Conclusions

We have presented a theoretical model of microparticle-assisted super-resolution microscopy and applied it to four representative structures commonly encountered in experiments: Blu-ray discs, slits, conducting dimers, and nanodisc arrays. For each geometry, appropriate illumination conditions were considered.

We demonstrated that the presence of rapid oscillations in the spatial correlation function of the radiation leads to a substantial enhancement of the resolving capability of the microparticle. Such oscillations may originate from different physical mechanisms. These include waves circulating inside the particle that generate secondary illumination of the sample [4], the specific geometry of the investigated object (e.g., a Blu-ray disc), and the choice of illumination conditions, as demonstrated here for annular illumination. In the latter case, the resolution was shown to depend on the size of the investigated structures, with larger objects exhibiting improved resolvability.

In the absence of rapid oscillations in the correlation function, super-resolution may still arise from a different mechanism: a local increase in the effective numerical aperture of the objective [5]. In this regime, the microsphere enables the discrimination of objects with center-to-center separations down to approximately $\lambda/2$.

**Funding.** Foundation for the Advancement of Theoretical Physics and Mathematics (23-1-1-61-2);

**Acknowledgements.** AB and BS thanks support from Foundation for the Development of Theoretical Physics and Mathematics «BASIS».

**Disclosures**. The authors declare no conflicts of interest.

**Data availability.** Data underlying the results presented in this paper are not publicly available at this time but may be obtained from the authors upon reasonable request.

Appendix

## 1. Source spectrum and geometry for DVD reflection calculations

In experiments, incoherent broadband light sources are typically employed. Fig. S1 shows the spectrum of the source used in the simulations (Fig. 4-5 of the main text). For the calculation of the image field, 10 discrete frequencies $\omega_i$ corresponding to specific wavelengths (indicated in the figure) were taken into account.

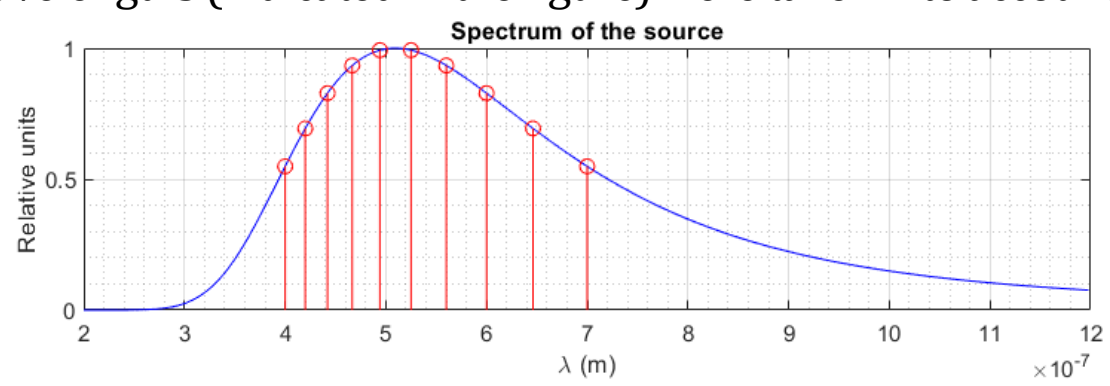

Fig. S1. Spectrum of the source used in the numerical simulations. In the image calculations, only 10 frequencies were considered, with the corresponding wavelengths indicated in the figure.

Figure S2 shows the detailed geometry of the simulation used to visualize the surface of a DVD disc with a microparticle immersed in water.

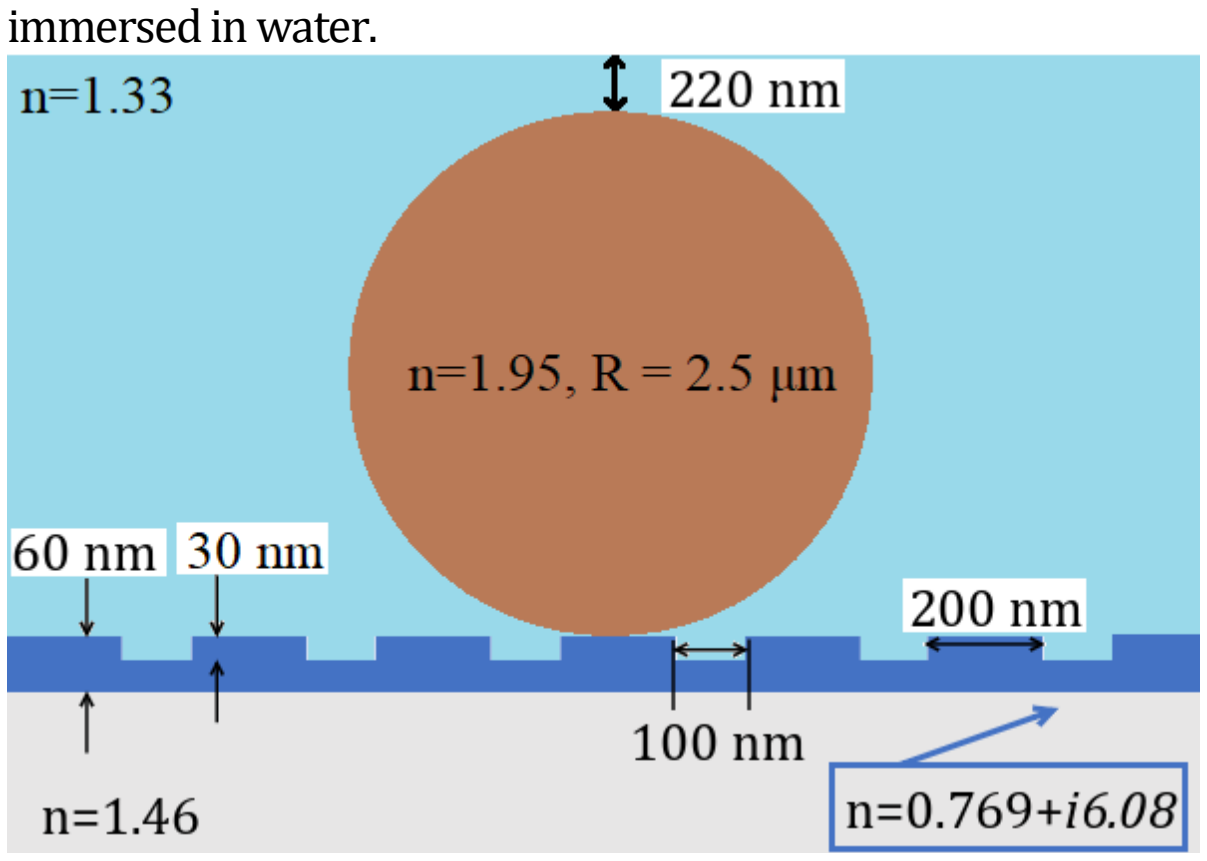

Fig. S2. Schematic view of the considered geometry. The immersion liquid, water (n = 1.33), barely covers the microparticle placed on the surface of the DVD disc.

## 2. Visualization of a dimer array under critical illumination

As an additional example, we considered the simulation of super-resolution imaging for an array of perfectly conducting dimers with a width of 120 nm and a center-to-center spacing of 60 nm, as reported in [9]. Figure S3(a) shows the geometry of the structure under study, while Fig. S3(b) presents the image obtained using the same method as employed for the result in Fig. 3(b). Although such aberration as distortion is seen in Fig. 3(b), center dimers are clearly distinguishable.

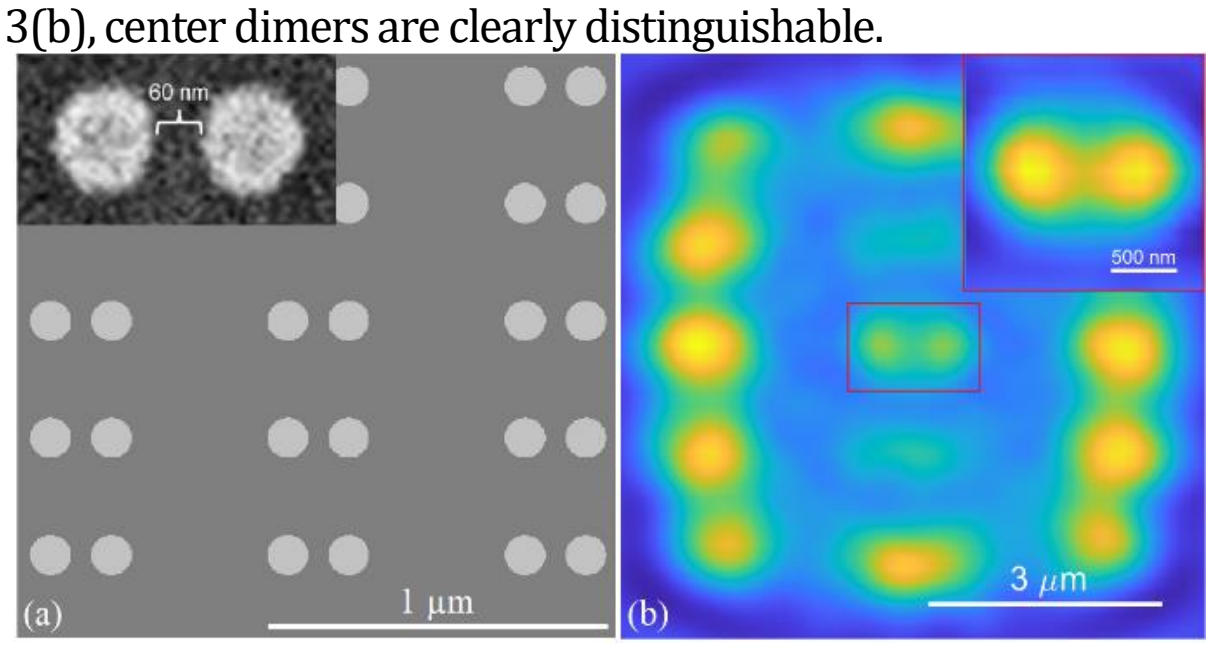

Fig. S3. (a) the geometry of the structure under study, (b) the image obtained using the same method as employed for the result in Fig. 3(b).

## 3. Influence of spatial coherence effects for small objects with and without the microsphere

In this section, we consider structures similar to those shown in Fig. 7 of the main text, but with the aperture diameters reduced to 40 nm (0.08 λ) while maintaining a center-to-center spacing of 180 nm. The results of the corresponding simulations are shown in Fig. S4(a–c). These results demonstrate that the structure becomes unresolved. Similarly, in free space without the microsphere, even when the aperture diameter is 200 nm, the objects remain unresolved (Fig. S4(d–f)).

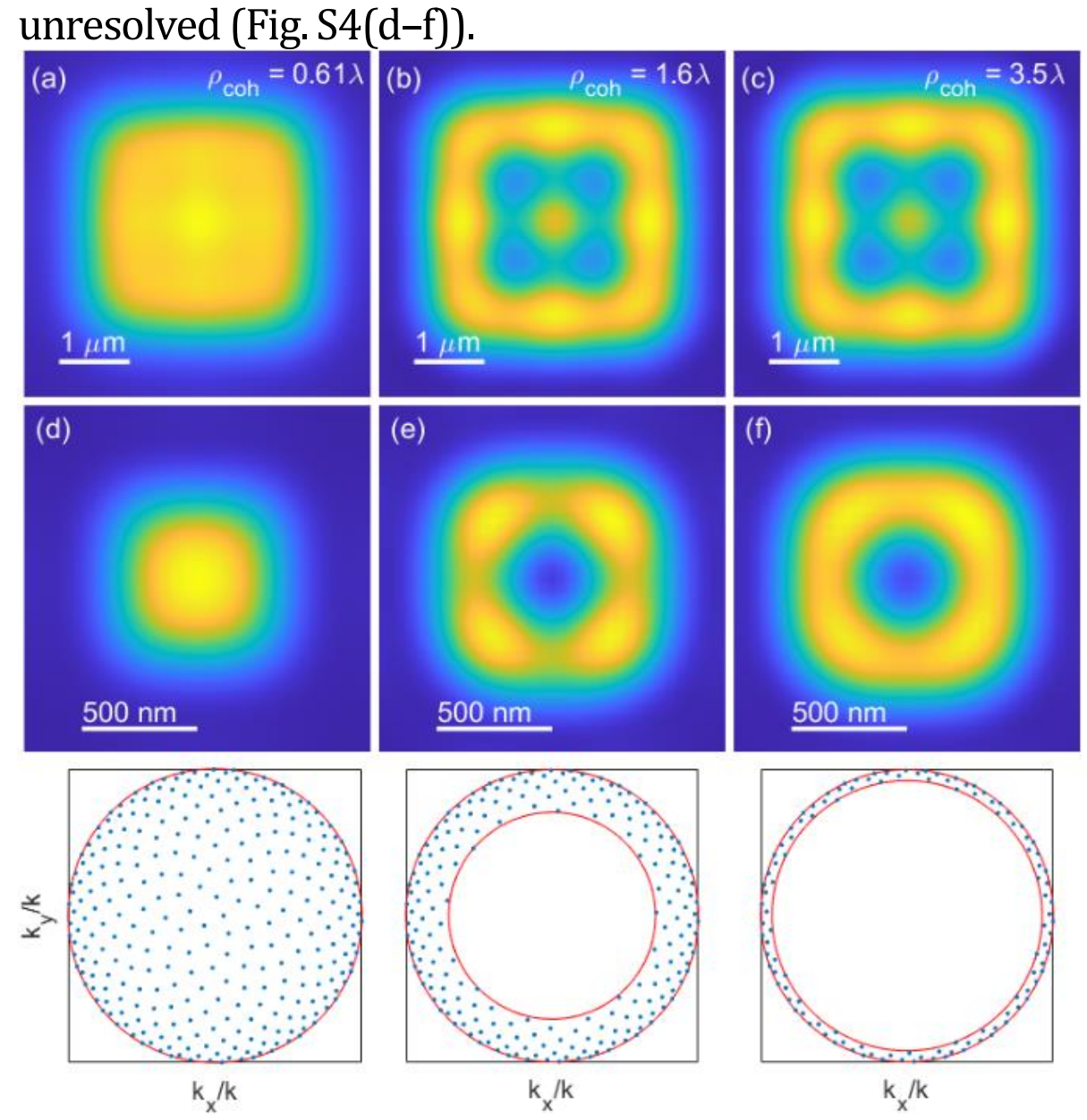

Fig. S4. Image field of an array of nine nanoscale apertures with a diameter of 40 nm and a period of 200 nm in a perfectly conducting film for different spatial coherence radii: (a–c) 0.61 λ, 1.6 λ, and 3.5 λ, respectively (λ = 500 nm). The image plane is at z/R = –5. Panels (d–f) show the corresponding images for apertures with a diameter of 180 nm in free space without the microsphere; the image plane coincides with the sample plane.

## 1. Image Calculation Algorithm Based on the Simple Model

The image calculation algorithm within the framework of the Huygens model differs depending on the imaging mode: transmission or reflection. A detailed technical description of each case is presented below.

*(A) Transmission Mode.* We first consider the simpler case of transmission. The image calculation in this mode involves the following steps:

*Step 1.* Source Definition: The source is positioned below the sample surface; in the examples considered here, a plane wave is used. Numerical simulations are performed in a volume encompassing the region near the contact point of the microsphere with the sample. The boundaries of the simulation domain must be located at a distance of at least 2–3 wavelengths from the edges of the sample, which is particularly important for oblique incidence.

*Step 2.* Field Formation on the Plane $\Gamma_{near}$: The propagation of radiation in the system is simulated using the standard Lumerical FDTD algorithm. The results are recorded as a Fourier representation of the field on the plane $\Gamma_{near}$ (see Fig.

2(c)–(d)). The size of this plane is not a critical parameter in the present case, but it must cover the region of slits or apertures in the mask.

*Step 3.* Application of the Huygens–Fresnel Principle: The field obtained on $\Gamma_{near}$ is then propagated using the Huygens–Fresnel principle, expressed as a diffraction integral:

$$\mathbf{E}^{near} = -\frac{k^2}{4\pi}\iint_{\Gamma_{near}} \begin{pmatrix} -i\frac{\mu_0}{n_0}[\mathbf{n},\mathbf{H}]G - i\frac{\mu_0}{n_0}\left([\mathbf{n},\mathbf{H}],\nabla\right)\nabla G \\ +\left[[\mathbf{n},\mathbf{E}],\nabla G\right] \end{pmatrix} dS \quad .(1)$$

Next, the diffracted field $\mathbf{E}^{near}$ on the microsphere is calculated using Mie theory. Since the field is known at a finite number of points on the simulation grid, the integral (1) reduces to a finite sum of point dipoles, for which the solution can be expressed analytically as a Mie series:

$$\mathbf{E}^{sca} = \sum_{l=1}^{\ell_{max}} \sum_{m=-l}^{l} a_{lm}\mathbf{N}_{lm}^{(sca)} + b_{lm}\mathbf{M}_{lm}^{(sca)} \ , \quad (2)$$

As the positions of the grid nodes remain unchanged for different illumination angles, this calculation needs to be performed only once for each point of the grid on $\Gamma_{near}$. For each illumination configuration, the corresponding sum of solutions is then taken, with amplitudes determined by the fields **E** and **H** in the integrand of Eq. (1).

*Step 4.* Field Calculation on the Plane $\Gamma$ Above the Microsphere: Finally, the total field is calculated on the plane $\Gamma$ above the microsphere (z=const>0):

$$\mathbf{E} = \mathbf{E}^{near} + \mathbf{E}^{sca} \ .$$

To accelerate the computations, $\mathbf{E}^{near}$ can also be expanded in a series as in Eq. (2). It is important to consider the convergence domain of the series: the distance from the sphere center to the plane $\Gamma$ must be greater than or equal to the distance from the sphere center to $\Gamma_{near}$; otherwise, the series diverges. The functions $\mathbf{M}_{lm}$ and $\mathbf{N}_{lm}$ entering Eq. (2) need to be computed only once for all illumination angles. The magnetic field **H** is calculated in a similar manner. Once **E** and **H** on $\Gamma$ are determined, the subsequent image calculation follows the same procedure as in the full model (FM).

*(B) Reflection Mode.* The reflection mode requires a more sophisticated approach due to the distortion of the wavefront caused by the microsphere. The calculation proceeds as follows:

*Step 1.* Source Definition: The diffraction of the incident plane wave $\mathbf{E}^{plane\ wave}$ on the spherical surface in free space is calculated. The resulting field distribution, both inside the microsphere and in the surrounding region, is used to define a source located near the contact point between the microsphere and the sample (see Fig. 2(g)). An *Imported Source* is used, which allows an arbitrary spatial field profile to be defined, taken from the Mie solution. The source domain must encompass both the field inside the particle and the surrounding field.

*Step 2.* Field Formation on the Plane $\Gamma_{near}$: This step is analogous to the transmission case, except that the size of $\Gamma_{near}$ must cover not only the sample region but also the focus of the field generated by the microsphere in the sample plane. For oblique incidence, the focal spot may be located at a significant distance from the point of contact. Reflected radiation from the substrate can contribute substantially to the resulting field; therefore, $\Gamma_{near}$ in reflection mode is generally larger than in transmission mode.

*Step 3.* Application of the Huygens–Fresnel Principle: This step is identical to the corresponding step in transmission mode.

*Step 4.* Field Calculation on the Plane $\Gamma$ Above the Microsphere: This step is also similar to the transmission case. However, in addition, the field scattered by the microsphere under the incident plane wave must be taken into account:

$$\mathbf{E} = \mathbf{E}^{near} + \mathbf{E}^{sca} + \mathbf{E}^{sca,\ plane\ wave} \ ,$$

where, $\mathbf{E}^{sca,\ plane\ wave}$ is the field scattered by the microsphere under illumination by $\mathbf{E}^{plane\ wave}$. Physically, this field does not contribute to the image formation; however, its inclusion is essential: $\mathbf{E}^{sca,\ plane\ wave}$ wave results in almost a twofold difference in the image field amplitude (Fig. 3 (d,e)).

It is important to note that fields on material surfaces may be discontinuous. In such cases, the "external" limit of the field should be considered, i.e., the field value approaching the surface from outside. In practice, this can be approximated by introducing a small gap between the plane $\Gamma$ and the object, equal to one simulation grid step. In the present calculations, this gap was 5 nm, meaning that the plane $\Gamma$ and the microsphere are positioned 5 nm above the conductor.